\def \cm{~\rm{cm}}
\def \s{~\rm{s}}
\def \km{~\rm{km}}

\def \AU{~\rm{AU}}

\def \yr{~\rm{yr}}

\documentclass[12pt,preprint]{aastex}
\usepackage{natbib}

\shorttitle{SHAPING THE RED RECTANGLE}
\shortauthors{Soker}
\slugcomment{Draft
version of \today}

\begin{document}

\title{THE SHAPING OF THE RED RECTANGLE PROTO-PLANETARY NEBULA}

\author{Noam Soker\altaffilmark{1}}

\altaffiltext{1}{Dept. of Physics, Technion, Haifa 32000, Israel;
soker@physics.technion.ac.il.}

\begin{abstract}
I argue that the slowly expanding biconical structure of the Red
Rectangle$-$a nebula around the post asymptotic giant branch
binary stellar system HD 44179$-$can be formed by intermittent
jets blown by the accreting companion.
The bright biconical structure of the Red Rectangle nebula
can be understood to be composed of a multiple double-ring system.
In the proposed shaping process, one among several processes through
which a companion can shape the circumbinary gas, the companion
accretes mass from the slow wind blown by the
evolved mass losing star.
An accretion disk is formed, and if mass accretion is
larger than a critical value, two jets, or a collimated fast
wind (CFW) are blown.
If the high mass loss rate duration is long, bipolar lobes are
formed.
If, on the other hand, mass loss rate is intermittent, and during
one orbital period the slow wind fills a region which does not extend
much beyond the binary system, then only a fraction of the double-lobe
structure is formed, namely, rings.
This, I propose, was the case with the progenitor of the Red Rectangle,
where intermittent episodes of enhanced mass loss rate led to the
formation of a multiple double-ring system.

\acknowledgements
\end{abstract}

\section{INTRODUCTION} \label{sec:intro}

It is commonly accepted now that planetary nebulae (PNs) and
similar nebulae around evolved stars, e.g., the nebula around the
massive star $\eta$ Carinae, owe their axisymmetrical shapes
to binary interaction.
There are many different processes by which a binary
companion can influence the mass loss geometry from an evolved star
(Soker 2004 and references there in).
Recent higher sensitivity surveys have found more PNs with binary
central systems,
(De Marco et al.\ 20004; Hillwig 2004; Pollaco 2004), further supporting
the binary paradigm.
In one of the shaping processes the companion accretes mass from the
mass losing primary star, with specific angular momentum high enough
to form an accretion disk and then blows two jets; if the jets are
not well collimated, they are termed collimated fast wind (CFW).
Detection of jets blown by the white dwarf (WD) companion
in symbiotic systems (Sokoloski 2004; Galloway \& Sokoloski 2004;
Brocksopp et al.\ 2004 and references therein) provides additional
support for the shaping by jets model.
Morris (1987) first proposed that many PNs are shaped by
jets blown by an accreting companion.
Other papers followed, attributing specific structural features
to jets:
Soker (1990) argues that dense blob pairs (termed ansae
or FLIERs) along the symmetry axis of many elliptical PNs are formed by
jets blown in the transition from the asymptotic giant branch (AGB)
to the PN phase (the proto-PN or post-AGB phase);
Sahai \& Trauger (1998) argue that jets blown in the post-AGB phase
shape many PNs;
In Soker (2002b) I demonstrated that jets can also account for the
presence of ring-pairs, as the two outer rings of SN 1987.
One of the most prominent structures formed by jets is a pair of lobes
(or bubble pair).
The similarity between the structure of some PNs and the structure of
X-ray deficient bubble-pairs in some clusters of galaxies leaves no
doubt that lobes in PNs are formed by fast jets (Soker 2003; 2004).
PNs whose main structure is two lobes with an equatorial waist between
them are termed bipolar-PNs.
Of course, not all PNs are shaped by jets.

Moreover, there are many processes by which binary companions
can influence the shape of the nebula, and there are many
evolutionary routes that progenitors of bipolar PNs can take
(Soker 2002c).
Considering the different processes, different evolutionary routes,
and the dependence of the appearance on inclination, I found
(Soker 2002c) that the number of different apparent structures
of bipolar PNs is larger than the number of known bipolar PNs and
proto-PNs.
Hence, I reasoned, every bipolar PN is a `unique' object in its
appearance, but all can be explained within the binary model paradigm.
Therefore, I suggest terminating the use of adjectives such as
`unique', `peculiar', and `unusual' to bipolar PNs and proto-PNs.
This is not just a terminology exercise.
By removing these adjectives the need to invoke a new model
for almost every `unusual' bipolar PN becomes unnecessary.
Instead, all bipolar PNs can be explained with a binary model,
using similar processes that occur in other PNs and/or other
astrophysical objects.

In a recent paper Cohen et al.\ (2004, here after CVBG; also Van Winckel 2004)
presented new images of the the Red Rectangle proto-PN, associated
with the post-AGB binary system HD 44179 (IR source AFGL 915).
This binary system has an orbital period of $T_{\rm orb} = 322$ days,
a semimajor axis of $a \sin i = 0.32 \AU$, and an eccentricity
of $e=0.34$ (Waelkens {\it et al.} 1996; Waters {\it et al.} 1998;
Men'shchikov et al.\ 2002).
CVBG attached to the nebula of the Red rectangle the
adjective `unique', while Van Winckel (2004), in reference
to my recommendation to stop using this adjective (Soker 2002c),
used the adjective `more unique', although he referred mainly to the
unique chemistry of the nebula (which might be indeed more
particularized than its morphology).
With a `unique' nebula, a `unique' model is required, and for that
CVBG cite the model proposed by  Icke (2004).
The simulations of Icke (2004) start with a periodic biconical flow
from the central region outward, based on the biconical
flow studied by Icke (1981).
My view is that the model proposed by Icke (2004) cannot account
for the structure of the Red Rectangle for the following reasons:
(1) The biconical outflow proposed by Icke (1981) is due to
a gaseous disk. In most astrophysical objects, however, from AGN to
young stellar objects, the accretion disk blows collimated winds,
or jets, but not biconical flows.
The large cold disk around the binary stellar system HD 44179 at
the center of the Red Rectangle (e.g., Jura \& Kahane 1999;
Dominik et al.\ 2003) is not expected to blow a wind energetic enough
to form the bipolar structure of the Red Rectangle.
(2) The simulations of Icke (2004) do not reproduce the exact structure
of the Red Rectangle.
In particular, the faint rims connecting the `vortices' to the center
(termed `winegalss (parabola)' in Fig. 10 of CVBG; reproduced here in
figure 1.) do not close back to the center in Icke's simulations.
Also, in the simulations the `ladder rungs' (Fig. 1)
are curved too much back toward the center.
(3) Most important, I am searching for a unified model to explain bipolar
PNs, based on a basic common driving agent. The model presented by
Icke (2004) seems to lack such an ingredient; the biconical outflow
is rare in other objects.

In the present paper I argue that jets blown by the accreting companion
can account for the structure of the Red Rectangle.
Using previously studied ingredients aiming at explaining bipolar nebulae,
from X-ray deficient bubbles pairs in cluster of galaxies through
supernovae remnants and PNs, I build a scenario to explain the basic
structure of the Red Rectangle.
In Section 2, I describe the basic structure of the Red Rectangle
and outline the proposed scenario.
In Section 3, I gives the relevant parameters which distinguish the
binary progenitor of the Red rectangle (HD 44179) from progenitors
having large orbital separation.
In Section 4, I summarize the main results.

\section{THE PROPOSED SCENARIO} \label{sec:scen}

\subsection{The structure of the Red Rectangle}

The general structure of the inner region of the Red Rectangle
is presented in Figure 10 of CVBG, which is reproduced here in Figure 1,
and depicts the following components.
There are two bright pairs of radial (more or less) lines, marked
`bicone edge' on Figure 1, emanating  from the center, each pair
on opposite sides of the equatorial plane, in an
approximate mirror symmetry.
This is the bright part of the biconical structure.
At larger distances, the opening angle of the biconical structure,
i.e., the angle between the rays, increases from $40^\circ$ to
$80^\circ$.
The bicone edge are composed of more or less symmetrical,
but not exactly periodic, bright knots, termed vortices by CVBG.
Faint arcs, in most cases not complete, connect the vortices to
the center, in a more or less parabolic shape; these are termed
`wineglass (parabola)' by CVBG.

It is assumed that the nebula has a more or less axisymmetrical structure,
and that the structure described above is a cut through a plane
containing the symmetry axis, and
that the symmetry axis of the Red Rectangle is almost parallel
to the plane of the sky (almost perpendicular to our line of sight).
Therefore, each pair of vortices on opposite side of the symmetry axis
are a limb-brightened projection of a single ring.
This ring is connected with a paraboloid to the center.
Such a structure is not unique to the Red Rectangle.
The outer lobe pair of the PN M2-9 has this structure (see HST home page
and Doyle et al.\ 2000).
In the outer lobes of M2-9 the paraboloid rims are brighter, the ring is
wider, and there is only one ring on each side of the equatorial plane.
However, the basic structure of a ring connected with faint paraboloid to the
center is similar to each ring in the Red Rectangle.

With somewhat less similarity, but still possessing this basic structure,
I list the Hourglass PN (MyCn 18), which also displays a wide bright ring
connected to the center with a paraboloid (Dayal et al.\ 2000).
Because the symmetry axis of MyCn 18 is highly inclined to the plane
of the sky, it is hard to compare it to the structure of the rings
in the Red Rectangle or to the rings structure in M2-9.

Sugerman et al.\ (2004) argue, based on their study of light echo
from SN 1987, that the outer double-ring structure of SN 1987
is formed by the bright parts of an hourglass structure connected
to the center of the nebula.
If this is indeed the case, then the outer rings of SN 1987
also have the ring-wineglass structure.

Double-ring systems have been found also in the PNs Hubble 12
(Welch et al.\ 1999) and He 2-113 (Sahai et al.\ 2000).
However, it is hard to tell whether the structure is that of
a ring-wineglass type.

\subsection{Intermittent jet launching}

From the previous subsection, I conclude that the basic difference between
the Red Rectangle and the other nebulae described, is that the progenitor
of the Red Rectangle (HD 44179) had multiple ejection events, each forming
a pair of ring-wineglass structure.

 The scenario is as follows. The AGB progenitors loses mass with episodes
of enhanced mass loss rate.
Semi-periodic enhanced mass loss rate episodes are inferred to occur in
stars about to leave the AGB from the presence of multiple semi-periodic
concentric shells (termed also arcs, rings, and M-arcs) in PNs
and proto PNs.
The M-arcs are observed mainly in the spherical haloes of elliptical
and bipolar PNs
(e.g., Hrivnak et al.\ 2001; Kwok et al.\ 2001; Corradi et al.\ 2004).
The companion accretes mass from the AGB star, forms an accretion
disk, and blow two jets, one at each side of the equatorial plane.
As I argued in a previous paper (Soker 2002b), such a process might
form a double-ring structure.
The mass loss rate between the enhanced mass loss rate episodes are too
low for the companion to blow jets.

The question I address now is why the Red Rectangle shows the
semi-periodic nebular structure in a biconical structure,
which I argued above is a multiple double-ring system,
while the other PNs mentioned above (e.g, Corradi et al.\ 2004)
show the multiple enhanced mass loss episodes
in more or less spherical structures?
The answer is in the accretion rate of the close companion in the
Red Rectangle, which is higher that in systems with larger
orbital separations.
In the Red Rectangle the orbital separation is $a \simeq 1\AU$,
while in the other PNs, according to the proposed scenario,
the orbital separation is larger.
Following the arguments presented by Soker \& Rappaport (2000),
I assume that jets are launched only when the accretion rate
into the companion is above a critical value of
$\dot M_{\rm crit} \sim 10^{-8}-10^{-7} M_\odot \yr^{-1}$,
depending on the nature of the companion, for example, a WD or
a main sequence star.

The accretion rate is the Bondi-Hoyle rate
$\dot M_2 = \rho_s \pi R_a^2 v_r$, where $\rho_s$ is the
density of the AGB wind, $R_a=2GM_2/v_r^2$ is the accretion radius,
$v_r$ is the relative velocity between the companion and the wind,
and $M_2$ is the mass of the compact companion.
The relative velocity is $v_r=(v_s^2+v_o^2)^{1/2}$, where $v_s$ is the
slow AGB wind velocity at the location of the companion, and $v_o$
is the orbital velocity of the companion around the primary.
It is adequate to assume here for the calculation of $v_o$ that
$M_1 \gg M_2$ and that the orbit is circular.
The wind density is $\rho_s = \dot M_1/(4 \pi a^2 v_s)$,
where $\dot M_1$ is the mass loss
rate of the AGB star, defined positively.
The accretion rate of the companion is given then by
\begin{equation}
\frac {\dot M_{2{\rm acc}}}{\dot M_1} =
\frac{v_o}{v_s}
\left( 1+ \frac {v_s^2}{v_o^2} \right)^{-3/2}
\left( \frac {M_2}{M_1} \right)^2
\equiv \Gamma_{\rm acc}
\left( \frac {M_2}{M_1} \right)^2,
\label{eq:acc01}
\end{equation}
where the accretion coefficient $\Gamma_{\rm acc}$ was
defined in the second equality.
To make the behavior more transparent, I take the slow wind speed to
be $v_s=10 \km \s^{-1}$, and $M_1=1 M_\odot$.
The expression for the accretion coefficient reads
\begin{equation}
\Gamma_{\rm acc} (v_s=10 \km \s^{-1}; M_1=1 M_\odot)=
\frac {3}{{a_u}^{1/2}}
\left( 1+ \frac {a_u}{9} \right)^{-3/2},
\label{eq:acc02}
\end{equation}
where $a_u$ is the orbital separation in units of $\AU$.

The first, and most important factor which determines the accretion
rate is the orbital separation.
For $a_u=1$, 5, and 25, one finds for $v_s=10 \km \s^{-1}$,
$\Gamma_{\rm acc}=2.56$, 0.69, and 0.08, respectively.
This demonstrates that the accretion rate is sensitive to the orbital
separation, a concept by now well established.

A second factor which enhances the accretion rate by companions
at very close separation is the wind speed.
The value of $\Gamma_{\rm acc}$ at an orbital separation of
$a \sim 1 \AU$ is likely to be even larger, since the wind there
does not reach its terminal velocity yet and $v_s$ is lower.

A third factor that increases the accretion rate of close companions
is tidal interaction.
Because of tidal interaction, a close companion is likely to enhance
the mass loss rate from the AGB star during the enhanced mass loss
rate episodes.

A fourth factor is the mass of the companion.
In order to avoid a common envelope phase, companions at small
orbital separations must be relatively massive, $M_s \ge 0.3 M_\odot$,
while companions at larger orbital separations, much larger than
the maximum envelope radius the primary attains on the AGB,
can be less massive.
Hence, in the Red Rectangle and similar objects the companion
cannot have too little mass.

\section{THE FLOW STRUCTURE} \label{sec:param}
\subsection{Jet Propagation}

In the model for double ring formation proposed in a previous paper
(Soker 2002b), the flow studied was axisymmetric, with two
jets expanding along the symmetry axis into a thin shell.
As stated earlier, such a flow is expected in binary systems where
the mass-losing primary undergoes an impulsive mass loss episode which
forms a thin, dense shell.
Each jet hits a substantial fraction of the shell's material.
The shocked shell's material is accelerated sideway by the jet,
and forms a higher density ring.
In that earlier paper I also noted that if there are several such
impulsive mass loss episodes, more double rings are formed, and,
because of the binary interaction and orbital motion, the double-ring
system is displaced from the symmetry axis of the main nebula.
Basically, the assumption in Soker (2002b) was that the orbital period
is much longer than the duration of the impulsive enhanced mass loss
episode.
The situation in the Red Rectangle is the opposite.
The enhanced mass loss rate episodes occur every 70 years
or longer (CVBG).
The duration of the enhanced mass loss episodes may last
some fraction of this, i.e., $\tau_{ep} \sim 10 \yr$.
The orbital period of the central binary system is
$\tau_o = 0.88 \yr \ll \tau_{ep}$.
In this section I discuss the implication of the very short orbital
period in regard to the formation of a double ring system during an
enhanced mass loss rate episode.

Let us consider the following flow structure, drawn schematically
in Figures 2 and 3.
The AGB (or post-AGB) wind flowing near the equatorial plane toward
the companion is accreted by the companion.
The wind above and below the equatorial plane and to the other side
of the companion is undisturbed.
This accretion flow from the primary to the companion
(not drawn on Figures 2 and 3) is as the flow simulated by
Mastrodemos \& Morris (1998, 1999).
The companion blows two jets, one on each side of the equatorial plane,
with the following properties.
The speed of gas inside the jet is $v_j$, its half opening angle
(from its symmetry axis to its edge) is $\theta$,
and the mass loss rate into each jet is
$\dot M_j = \beta \dot M_{2{\rm acc}}$, where
$\beta$ is the fraction of the accreted mass blown into each jet.
In driving the approximate equations below I take $\theta \ll 1$,
and $v_j \gg v_s$.
The density inside the jet, which propagates perpendicular to
the orbital plane along the $z$ axis, is
\begin{equation}
\rho_j =  \frac {\beta \dot M_{2{\rm acc}}} {\pi z^2 \theta^2 v_j}.
\label{eq:rhoj}
\end{equation}
The density of the slow wind material, through which the jet
plows, at height $z$ from the equatorial plane is given by
\begin{equation}
\rho_s=\frac {\dot M_1}{4 \pi v_s} \frac{1}{a^2+z^2},
\label{eq:rhos}
\end{equation}
where $a$ is the orbital separation.
The jets interacts with the slow wind (before breaking out), and material
flows to the sides of the jet's heat, marked as Region A in Figure 2 and
Region B in Figure 3.

The head of the jet proceeds at a speed $v_h$ given by the balance
of pressures on its two sides.
Assuming supersonic motion and $v_h \gg v_s$, this equality reads
$\rho_s v_h^2 = \rho_j (v_j-v_h)^2$.
Eliminating $v_j/v_h$, using the equations above for $\rho_j$,
and $\rho_s$, and with $\dot M_{2{\rm acc}}$ from
equations ({\ref{eq:acc01}}) and ({\ref{eq:acc02}}),
yields
\begin{equation}
\frac {v_j}{v_h}-1 \simeq
\frac {z}{(a^2+z^2)^{1/2}} \frac{\theta}{2 (\Gamma_{\rm acc} \beta)^{1/2}}
\left( \frac{v_j}{v_s} \right)^{1/2}
\frac {M_1}{M_2}.
\label{eq:vh1}
\end{equation}
Close to the jet's source, $z \ll a$, the jet's head proceed
at a speed close to $v_j$.

The jets clean a region above the companion.
When the companion returns to that point after one revolution,
the slow wind filled that region up to a distance from the
AGB star equals to $r_s=R_1+v_s \tau_o$.
the outflow in the Red Rectangle has $v_s<10 \km \s^{-1}$
(Jura et al.\ 1997), and as in CVBG I use $v_s=7 \km \s^{-1}$.
For $R_1=0.5 \AU$ and $\tau_o=0.88 \yr$, I find $r_s=1.8 \AU$.
For $a =1 \AU$, then, in order to break out of the
slow wind dense region the jets should reach a distance
of $z \simeq (r_s^2-a^2)^{1/2} \simeq 1.5 \AU$.
However, as the jet's head proceeds, the companion moves a substantial
distance, such that no fresh jet material is supplied.
The material pushed to the side along the orbit (Region B in Fig. 3),
is later accelerated along the $z$ direction, because of the companion
orbital motion.  This accelerated gas is drawn schematically as
Region C in Figure 3.
Later the jet's head breaks out of the slow wind gas (because the
primary mass-losing star did not fill this region), and pushes
the material to form the ring (drawn as region D in Fig. 3).

To examine the jet's head motion, I use in equation ({\ref{eq:vh1}})
the typical values $M_2=0.5 M_1$, $V_j=300 \km \s^{-1}$,
$\theta=0.2$ (opening angle of $\sim 10 ^\circ$),
$\beta=0.05$ ($10\%$ per cents of the accreted mass is blown into
the two jets), and $\Gamma_{\rm acc}=2$, resulting in $v_h=0.25v_j$
at $z=a \simeq 1 \AU$.
I neglect the term $`-1'$ in the left hand side of equation
({\ref{eq:vh1}}), hence overestimating somewhat the speed of the jet's
head $v_h$ at $z \sim 1 \AU$, and obtain
\begin{equation}
\frac {v_h}{v_o} \lesssim
\frac {(a^2+z^2)^{1/2}}{z} \frac{2 (\Gamma_{\rm acc} \beta)^{1/2}}{\theta}
\left( \frac{v_s}{v_j} \right)^{1/2}
\frac{v_j}{v_o}
\frac {M_2}{ M_1}.
\label{eq:vh2}
\end{equation}
The total width of the jet at a distance $z$ from the equatorial
plane is $2 \theta z$ (see Fig. 3).
The orbiting jet-blowing companion crosses this distance along
its orbital motion in a time of $t_c=2 \theta z /v_o$.
During that time the jets proceeds a distance of
\begin{equation}
\Delta z \simeq t_c v_h \simeq
4 (\Gamma_{\rm acc} \beta)^{1/2} (a^2+z^2)^{1/2}
\left( \frac{v_s}{v_j} \right)^{1/2}
\frac{v_j}{v_o}
\frac {M_2}{ M_1}.
\label{eq:deltaz}
\end{equation}
Note that the distance does not depend on the opening angle of the jet.
For the parameters used in the previous example, and $v_o=30 \km \s^{-1}$
one finds $\Delta z \sim (a^2+z^2)^{1/2} > z$ and for
$z \simeq a$,  $\Delta z \sim a \sim 1 \AU$.
For these parameters, therefore, the jets break out of the slow wind
at about the same time that no fresh jet's material is supplied to
accelerate its head.
The jets push the slow wind dense gas to the sides
(Regions C and D in Fig. 3), forming rings.
The slow wind gas which is pushed to the side in the radial direction
(Region A in Fig. 2), will form the bright rim on the wineglass
structure.

It was assumed here that the two jets propagate perpendicular to the
orbital plane.
This is a good approximation.
Jet bending by the ram pressure of the slow wind is small for the
typical parameters used here.
This can be shown by using equation (11) of Soker \& Rappaport
(2000), which shows that significant bending occurs only
at distances of $z \gg 1 \AU$, which requires larger
orbital separation than that in the Red Rectangle.
A small departure from perpendicular expansion relative
to the center of mass of the system occurs because of the
orbital motion. The pre-shock jet's material has a velocity
component parallel to the orbital plane of $v_o$, which is, however,
small relative to the perpendicular velocity of $v_j$.

\subsection{Rings Velocity}

To estimate the ring velocity, I first note that at the
distances relevant to the flow here, the densities are high
and all gas segments involved cool rapidly.
Hence, momentum conservation is a good approximation.
Let the ring material be composed of slow wind material that
was blown inside a cone having a half opening angle of $\phi_s$,
measured from an axis originating from the primary
mass losing star and perpendicular to the orbital plane.
A wind segment moving at angle $\phi$ has a velocity
$v_p=v_s \cos \phi$ perpendicular to the orbital plane,
and $v_t= v_s \sin \phi$ parallel to the orbital plane.
Integrating the absolute value of the momentum of slow wind segments
will give the momentum discharge (momentum per unit time)
component perpendicular and parallel to the orbital plane.
The mass flow rate into one cone is
$\dot M_{1c}=\dot M_1 (1-\cos \phi)/2$.
The total momentum discharge perpendicular to the orbital plane in
one cone is
\begin{equation}
\dot p_{sz}=\dot M_{1c} v_s (1-\cos \phi_s)^{-1}
\int_0^{\phi_s} \cos \phi \sin \phi d \phi =
\dot M_1 v_s \frac {\sin^2 \phi_s}{4},
\label{eq:pcz}
\end{equation}
while the total momentum flow rate parallel
to the orbital plane in one cone is
\begin{equation}
\dot p_{sp}=\dot M_{1c} v_s (1-\cos \phi_s)^{-1}
\int_0^{\phi_s} \sin^2 \phi d \phi =
\dot M_1 v_s \frac {2 \phi_s - \sin 2 \phi_s}{8}.
\label{eq:pcp}
\end{equation}
The jet blown by the companion adds a momentum mainly perpendicular
to the orbital plane at a rate of
\begin{equation}
\dot p_{jz}=\beta \dot M_{2{\rm acc}} v_j=
\beta \dot M_1 v_j \Gamma_{\rm acc} \left( \frac {M_2}{M_1} \right)^2
\label{eq:jz}
\end{equation}

For the parameters used here ($\beta=0.05$; $\Gamma_{\rm acc}=2$;
$v_j/v_s=300/7$; $M_2/M_1=0.3$), I find for $\phi_s=60^\circ$, that
$p_{sz}=  0.19 \dot M_1 v_s$, $p_{sp}= 0.15 \dot M_1 v_s$, and
$p_{jz}= 0.39 \dot M_1 v_s$.
If all the jet's momentum is deposited to the ring, then the
half-opening angle of the rings will be
$\tan \psi_{\rm ring} = p_{sp}/(p_{sz}+p_{jz})$, or
$\psi_{\rm ring}= 15^\circ$.
For $\phi_s=70^\circ$, I find
$p_{sz}=  0.22 \dot M_1 v_s$ and $p_{sp}= 0.23 \dot M_1 v_s$
($p_{jz}$ does not change), and $\psi_{\rm ring}= 20^\circ$.
In the Red Rectangle (see  CVBG) the half-opening angle of the
rings increases from $20^\circ$ close to the center to $30^\circ$
in the region shown in Figure 1, to $40^\circ$ in the outer nebula.
In the present proposed model for the multiple-ring structure, this
increase of the opening angle as rings move away from the center
is explained as transverse acceleration at later times.
As shown in section 3.1, the jets are likely to break out of the
dense, slow-wind material blown during the enhanced mass loss episode.
However, between these enhanced mass loss episodes the more
diluted  slow-wind gas continues to fill the region up to the ring
(assuming no jets are blown between enhanced mass loss episodes).
Before the next enhanced mass loss rate episode, the ring reaches
a distance of $z \gtrsim 10^{15} \cm$.
The jets themselves, after breaking out of the denser slow wind material
blown during an enhanced mass loss rate episode,
will fill regions further out.
At these large distances the cooling time becomes longer than the flow
time, and a hot bubble might be formed, that accelerates mass to the
sides, as in the model for bubble formation by jets
(Soker 2002a; Lee \& Sahai 2003).

\section{DISCUSSION AND SUMMARY} \label{sec:summary}

The central goal of the paper is to show that the main structure of
the Red Rectangle (central star HD 44179; IR source AFGL 915)
does not need a unique model; it can be explained
by the same process that was described in forming double-ring systems
(Soker 2002b) and lobes in planetary nebulae and related objects,
e.g. the outer rings in SN 1987a.
Namely, a companion accretes mass from the evolved mass losing star
(the primary). If the specific angular momentum of the accreted mass
is high enough an accretion disk is formed, and if, in addition, mass
accretion rate is high enough jets (or a collimated fast wind; CFW)
are blown.
In this CFW model, if the mass loss rate is more or less continuous,
the CFW (or jets) lead to the formation of double lobes
(or double bubbles) with a waist between them, i.e.
a bipolar structure.
This model can account for bipolar structure in symbiotic nebula, PNs,
the nebula around the massive star $\eta$ Carinae, and other
related objects.
If, on the other hand, mass loss rate is intermittent, and during
one orbital period the slow wind fills a region which does not extend
much beyond the binary system, then only a fraction of the double-lobe
structure is formed, namely, rings.
In the Red Rectangle this is the case, as the slow wind velocity is
much lower than the orbital motion, $v_s \sim 0.25 v_o$, and during one
orbital period the slow wind fills a region comparable to the binary separation.

This work is motivated by the recent detail study of the structure of
the Red Rectangle conducted by CVBG.
As discussed in section 2.1, I interpret the bright structure to be
composed of several ring pairs (Fig. 1 taken from figure 10 of CVBG).
In Section 2.1, I presented the basic scenario of semi-periodic
enhanced mass loss rate episodes. These lead to the formation of jets
by the accreting companion.
As was shown in a previous paper (Soker 2002b) such jets can form rings.
The difference between the Red Rectangle and many other binary systems
that form bipolar PNs is the small orbital separation which enable jet
formation at relatively low mass loss rate (still, an enhanced mass loss
rate episode is required), and the slow wind speed being much lower than
the orbital velocity (Section 2.2).
In Section 3.1, I found that the jets breaks out of the region
filled by the dense slow wind during one orbital period, but they
marginally do so.
The material pushed to the side and forward in one enhanced mass
loss rate episode form one ring (Figs. 1 and 2).
In Section (3.2) I showed that the rings will move at a relatively
low speed when compared with the jets' speed.

The present study is by no means a substitute to 3 dimensional
hydrodynamical gas simulation.
However, such numerical simulations are resource demanding, and the
basic flow structure should be explored via analytical means
to make the numerical simulations as realistic as possible.

As shown in previous works (e.g., Soker \& Rappaport 2001),
in the binary shaping model of nebulae around evolved stars,
if the orbital orbit is eccentric the nebula is expected to possess
departure from pure axisymmetrical structure (binary progenitor in
circular orbit can also lead to departure from axisymmetry).
Such a departure is observed indeed in the Red Rectangle,
in the equatorial plane in the near infrared contour map of
Tuthill et al.\ (2002, Fig. 1), and in the infrared maps
of Miyata et al.\ (2004, Fig. 1).
(Analyzing the departure from axisymmetry is an effective tool to
understand the shaping of nebulae, which, oddly is
ignored by most observers.)

The conclusion of this paper is that the slowly expanding biconical
structure of the Red Rectangle might be formed by intermittent jets
blown by an accreting companion, thereby demonstrating the suitability
of the Red Rectangle in the binary shaping model of PNs.

\acknowledgements
I thank Ann Feild from the Space Telescope Science Institute for
providing Figure 1, and Martin Cohen for his permission to use
this figure.
This research was supported in part by the Israel Science Foundation.


\begin{figure}
\caption{Schematic drawing of the Red
Rectangle nebula as presented in Figure 10 of CVBG. The new
structural elements discovered by CVBG are marked, with one
addition: I identified their `vortices' with rings.  }
\end{figure}

\begin{figure}
\caption{The schematic flow structure near
the binary system during an enhanced mass loss rate episode. A
fraction of the slow wind blown by the AGB (or post AGB) star,
marked $M_1$, at a speed of $v_s \simeq 7-10 \km \s^{-1}$ is
accreted by the companion via an accretion disk (eq.
{\ref{eq:acc01}}). Two jets are assumed to be blown along the $z$
axis. The gas inside the jets moves at a speed
$v_j \sim 300 \km \s^{-1} \gg v_s$, and interacts with the slow
wind in shock waves.
The jet's head proceeds at a speed of $v_h$ (eq. {\ref{eq:vh1}}).
Post shock slow wind gas is pushed to the sides; a small portion
of this gas, on one side of the jet, is drawn schematicaly as
Region A. The orbital motion is perpendicular to the plane of the figure.
For clarity, some features are shown only on one side of the
equatorial plane. }
\end{figure}

\begin{figure}
\caption{Same as Figure 2, but in a plane momentarily tangent to the
secondary orbital trajectory.
The primary mass-losing star (not shown) is behind the companion;
$v_o \simeq 30 \km \s^{-1}$ is the companion's orbital velocity.
$\Delta z$ is the distance the gas inside the jet moves
at a distance $z$ from the companion, during the time the companion moves
a distance equal to the widths of the jet at $z$: $2 \theta z$
(eq. {\ref{eq:deltaz}}).
During one orbital period, the slow wind gas fills the region
up to a distance marked by a dotted line.
The post-shock slow wind material is drawn schematically as Regions
B-D. As the companion moves, material presently on Region B will
be accelerated outward to Regions C and D.
This dense gas will form one ring; another ring is on the other side
of the equatorial plane. Intermittent enhanced mass loss rate episodes
will form a multiple-ring structure.  }
\end{figure}


\begin{references}

\reference{} Brocksopp, C., Sokoloski, J. L., Kaiser, C., Richards, A. M.,
      Muxlow, T. W. B., Seymour, N. 2004, MNRAS, 347, 430

\reference{} Cohen, M., Van Winckel, H., Bond, H. E., \& Gull, T. R.
  2004, AJ, 127, 2362 (CVBG)

\reference{} Corradi, R. L. M., Sanchez-Blazquez, P., Mellema, G.,
     Giammanco, C., \& Schwarz, H. E. 2004, A\&A, 417, 637

\reference{}  Dayal, A., Sahai, R., Watson, A. M., Trauger, J. T., Burrows, C. J.,
    Stapelfeldt, K. R., \& Gallagher, J. S., III 2000, AJ, 119, 315

\reference{} De Marco, O., Bond, H. E., Harmer, D., Fleming, A. J.
 2004, ApJ, 602, L93

\reference{} Dominik, C., Dullemond, C. P., Cami, J., \&
  van Winckel, H. 2003, A\&A, 397, 595

\reference{} Doyle, S., Balick, B., Corradi, R. L. M., \& Schwarz, H. E.
     2000, AJ, 119, 1339

\reference{} Galloway  D. K., \& Sokoloski, J. L. 2004, ApJ, 613, L61


\reference{} Hillwig, T.\
  2004, in Asymmetrical Planetary Nebulae III: Winds, Structure and
  the Thunderbird, eds. M. Meixner, J. H. Kastner, B. Balick,
  \& N. Soker, ASP Conf. Series, 313, (ASP, San Francisco)
   (astro-ph/0310043)

\reference{} Hrivnak, B.\ J., Kwok, S., \& Su, K.\ Y.\ L.\ 2001,
     AJ, 121, 2775

\reference{} Icke, V. 1981, ApJ 247, 152

\reference{} Icke, V. 2004,
http://www.strw.leidenuniv.nl/
("The riddle of the REd Rectangle")

\reference{} Jura M., \& Kahane C., 1999, ApJ, 521, 302

\reference{} Jura M., Turner, J., \& Balm, S. P. 1997, ApJ, 474, 741

\reference{} Kwok, S., Su, K. Y. L., \& Stoesz, J. A. 2001,
in Astrophysics and Space Science Library Vol. 265, Post-AGB Objects
as a Phase of Stellar Evolution, ed. R. Szczerba \& S.K.Gorny
(Dordrecht: Kluwer), 115

\reference{} Lee, C.-F., \&  Sahai, R. 2003, ApJ, 586, 319

\reference{} Mastrodemos, N., \& Morris, M. 1998, ApJ, 497,303

\reference{} Mastrodemos, N., \& Morris, M. 1999, ApJ, 523, 357

\reference{} Men'shchikov, A. B., Schertl, D., Tuthill, P. G.,
   Weigelt, G., \& Yungelson, L. R. 2002, A\&A, 393, 867

\reference{} Miyata, T., Kataza, H., Okamoto, Y. K., Onaka, T.,
   Sako, S., Honda, M., Yamashita, T., \& Murakawa, K. 2004, A\&A
    415, 179

\reference{} Morris, M. 1987, {PASP}, 99, 1115

\reference{} Pollacco, D.\
  2004, in Asymmetrical Planetary Nebulae III: Winds, Structure and
  the Thunderbird, eds. M. Meixner, J. H. Kastner, B. Balick,
  \& N. Soker, ASP Conf. Series, 313, (ASP, San Francisco)

\reference{} Sahai, R., Nyman, L.-A., Wootten, A. 2000, ApJ, 543, 880

\reference{} Sahai, R., \& Trauger, J. T. 1998, AJ, 116, 1357

\reference{} Soker, N. 1990,  AJ, 99, 1869

\reference{} Soker, N. 2002a, ApJ, 568, 726  

\reference{} Soker, N. 2002b, ApJ, 577, 839  

\reference{} Soker, N. 2002c, MNRAS, 330, 481 

\reference{} Soker, N. 2003, PASP, 115, 1296 

\reference{} Soker, N. 2004, in Asymmetrical Planetary Nebulae
III: Winds, Structure and the Thunderbird, eds. M. Meixner, J. H. Kastner, B. Balick,
\& N. Soker, ASP Conf. Series, 313, (ASP, San Francisco)
(extended version on astro-ph/0309228)

\reference{} Soker, N.  \& Rappaport, S. 2000, ApJ, 538, 241

\reference{} Soker, N.  \& Rappaport, S. 2001, ApJ, 557, 256

\reference{} Sokoloski, J. L. 2004, to appear in the Journal of the American
 Association of Variable Star Observers (AAVSO) (astro-ph/0403004)

\reference{} Sugerman, B. E. K., Crotts, A., Lawrence, S., Kunkle, B.,
  \& Heathcote, S. 2004,
  in Asymmetrical Planetary Nebulae
  III: Winds, Structure and the Thunderbird, eds. M. Meixner, J. H. Kastner, B. Balick,
  \& N. Soker, ASP Conf. Series, 313, (ASP, San Francisco)
  (extended version on astro-ph/0309228)

\reference{}  Tuthill, P. G., Men'shchikov, A. B., Schertl, D.,
  Monnier, J. D., Danchi, W. C., \& Weigelt, G.  2002, A\&A, 389, 889

\reference{} Van Winckel, H.
  2004, in Asymmetrical Planetary Nebulae III: Winds, Structure and
  the Thunderbird, eds. M. Meixner, J. H. Kastner, B. Balick,
  \& N. Soker, ASP Conf. Series, 313, (ASP, San Francisco) (see:
\newline
http://www.astro.washington.edu/balick/APN/APN\_talks\_posters.html
\newline
and go to H. Van Winckel talk).

\reference{} Waelkens, C., Van Winckel, H., Waters, L. B. F. M.,
\& Bakker, E. J. 1996, A\&A, 314, L17.

\reference{} Waters, L. B. F. M. {\it et al.} 1998, Nature, 391, 868

\reference{} Welch, C. A.,  Frank, A., Pipher, J. L., Forrest, W. J.,
    \& Woodward, C. E. 1999, ApJ, 522, L69


\end{references}
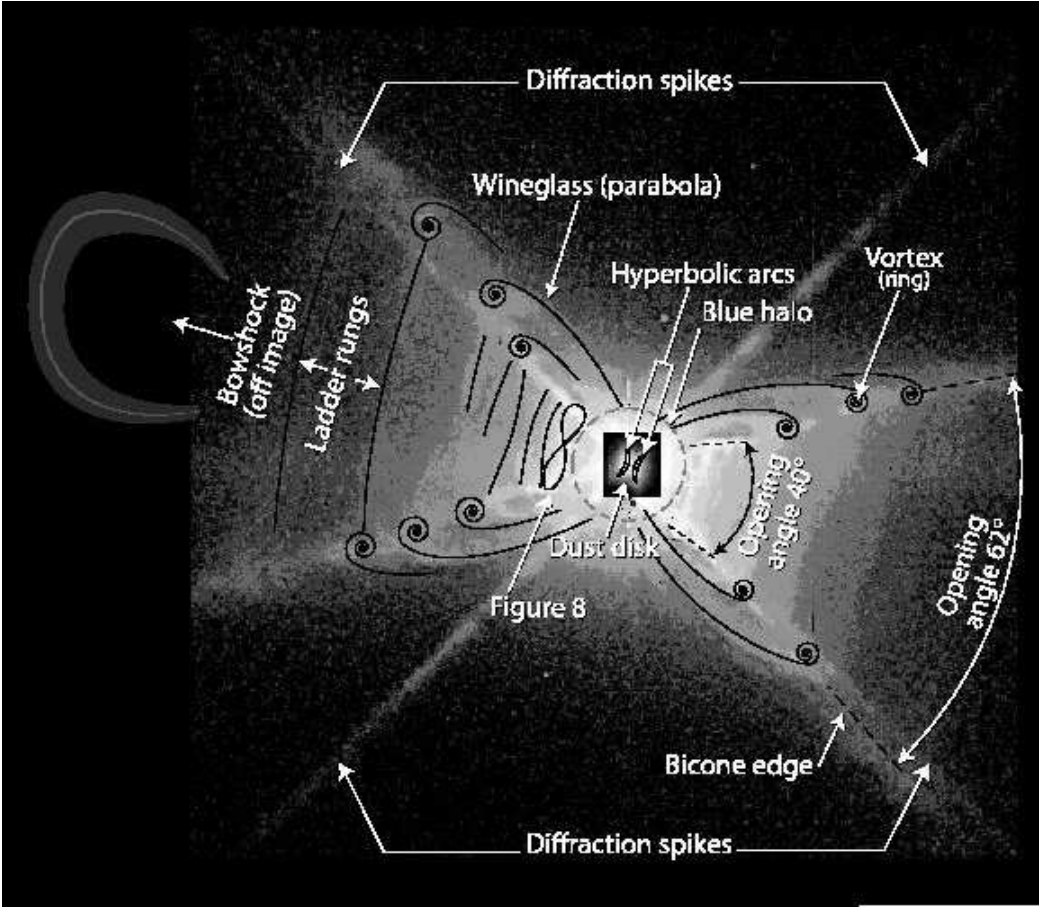
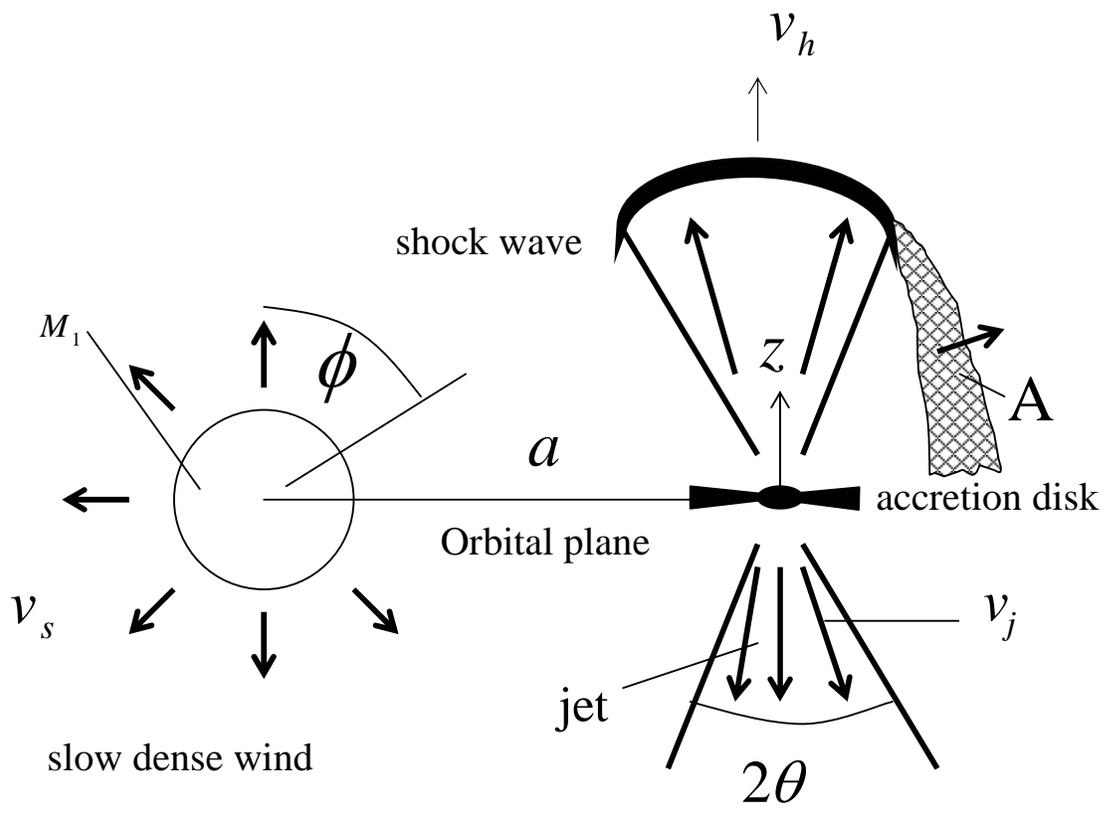
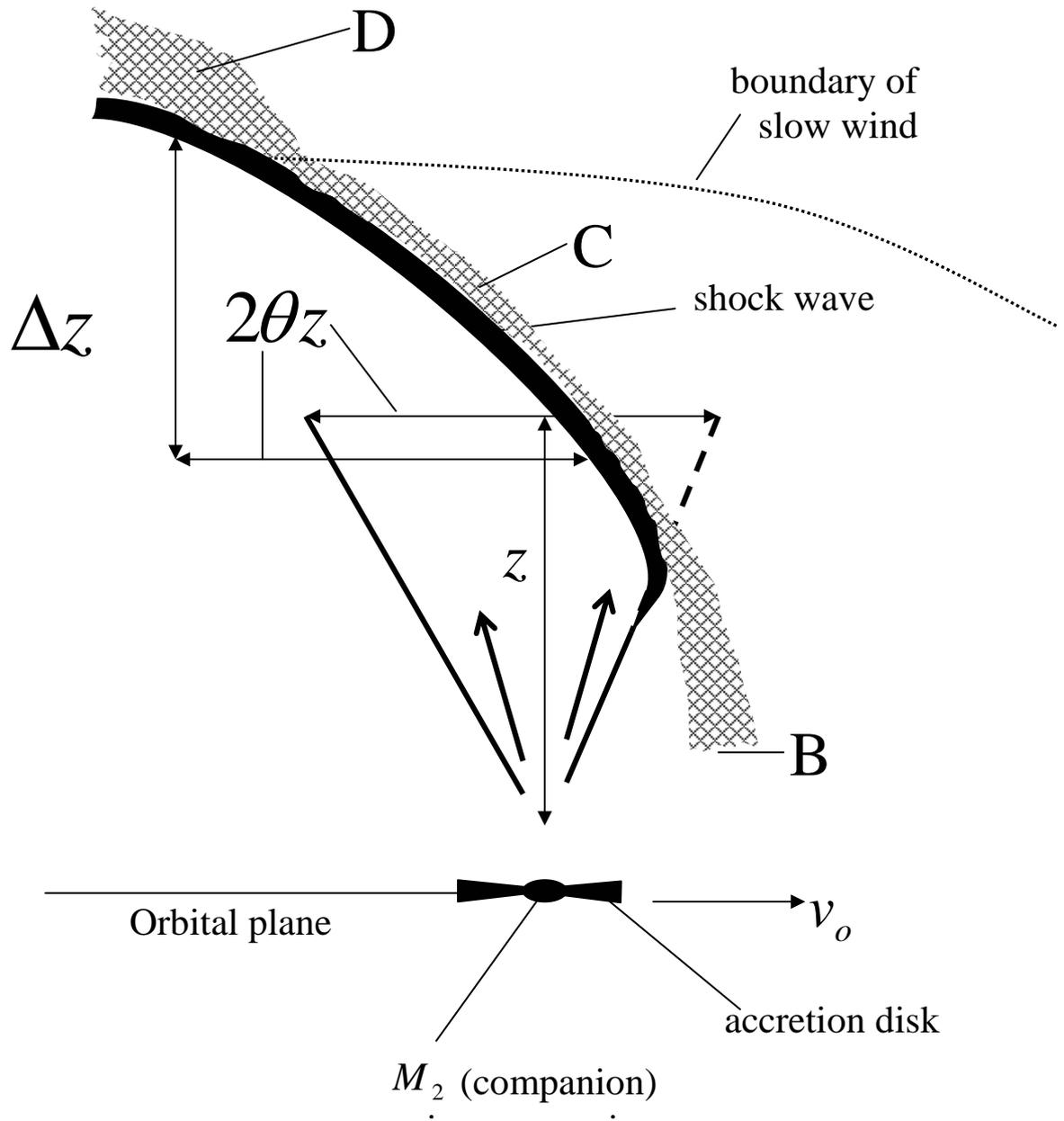
\end{document}